\documentclass[aps,prl,reprint,groupedaddress,float]{revtex4-2}

\usepackage{natbib,graphicx}
\usepackage{siunitx,amsmath} 
\usepackage{color}
\newcommand{\Alfven}{Alfv\'{e}n }

\newcommand{\V}[1]{\mathbf{#1}}

\newcommand{\vhat}[1]{\ensuremath{\hat{\mathbf{#1}}}}

\newcommand{\blue}[1]{{#1}}

\begin{document}


\title{MMS Observations of the Velocity-Space Signature of Shock-Drift Acceleration}


\author{P.~Montag}
\email[]{peter-montag@uiowa.edu}
\author{G.~G.~Howes}
\author{D.~McGinnis}
\author{A.~S.~Afshari}
\affiliation{Department of Physics and Astronomy, University of Iowa, Iowa City, Iowa 52242, USA.}
\author{M.~J.~Starkey}
\author{M.~I.~Desai}
\affiliation{Southwest Research Institute, San Antonio, TX, USA.}


\date{\today}

\begin{abstract}
Collisionless shocks play a key role in the heliosphere at planetary bow shocks by governing the conversion of the upstream bulk kinetic energy of the solar wind flow to other forms of energy in the downstream, including bulk plasma heat, acceleration of particles, and magnetic energy.  Here we present the first observational identification of the velocity-space signature of shock-drift acceleration of ions at a perpendicular collisionless shock, previously predicted using kinetic numerical simulations, using a field-particle correlation analysis of Magnetospheric Multiscale (\emph{MMS}) observations of Earth's bow shock.  Furthermore, by resolving the ion energization rates as a function of particle velocity, the field-particle correlation technique facilitates a clean quantitative separation of the energization rate of the reflected ions from that of the incoming ion beam, enabling a more complete characterization of the energy conversion at the shock.
\end{abstract}


\maketitle


\emph{Introduction.}---Collisionless shocks govern the energy transport occurring at planetary bow shocks, transitioning the upstream supersonic and super-Alfv\'enic flow into a downstream subsonic flow.  The dynamics within the shock transition taps the kinetic energy of the upstream flow both to heat the downstream plasma and to accelerate a small fraction of particles to higher energy.  For supercritical shocks, resistivity is insufficient to provide the dissipation needed by the Rankine-Hugoniot shock jump conditions  \cite{Edmiston:1984}, leading to the reflection back upstream of a fraction of the incoming ions and their acceleration by the upstream motional electric field, a process known as \emph{shock-drift acceleration} (SDA) \cite{Paschmann:1982,Sckopke:1983,Anagnostopoulos:1994,Ball:2001}. 

The nonthermal acceleration of particles at collisionless shocks is an inherently kinetic mechanism, calling for the examination of the energization of different populations of particles (\emph{e.g.}, incoming vs.~reflected ions) in the 3D-3V phase space of kinetic plasma theory.  The field-particle correlation technique \citep{Klein:2016a,Howes:2017a,Klein:2017b} was devised as a means of probing the energization of particles in weakly collisional plasmas using single-point measurements of the electromagnetic fields and particle velocity distribution functions.  It has been applied to explore the energization of ions and electrons in 1D-2V kinetic simulations of a perpendicular collisionless shock \citep{Juno:2021}, determining the unique velocity-space signatures of shock-drift acceleration of ions and adiabatic heating of electrons.  More recently, it has been applied to quasiperpendicular shock simulations with a shock normal angle of $\theta_{Bn}=\qty{45}{\degree}$, showing that particles that reflect off the shock multiple times can be clearly separated in velocity space \cite{Juno:2022} and that a variation of the field-particle correlation technique can be used to isolate the energization of particles due to kinetic instabilities arising in the shock transition \cite{Brown:2022}.

Here we report on the first successful application of the field-particle correlation technique to identify the velocity-space signature of shock-drift acceleration at a perpendicular collisionless shock using observations from the \emph{Magnetospheric Multiscale} (\emph{MMS}) mission \cite{Burch:2016} at Earth's bow shock.  Using the detail provided by the full velocity-space correlation, this technique makes possible the separation of the energization of the incoming ions from that of the reflected ions, enabling a more complete characterization of the overall energy transport and the nonthermal particle acceleration at the shock transition.

\emph{Data and Analysis Techniques.}---We analyze data from the MMS3 spacecraft on 2018-11-20 in the time interval 06:10:53-06:13:42 UT, where a bow shock crossing is observed at 06:12:04. Magnetic field data were measured by Fluxgate Magnetometers (FGM) \cite{Russell:2016}, electric field data by Electric field Double Probe (EDP) \cite{Ergun:2016,Lindqvist:2016}, particle moment and velocity distribution data by Fast Plasma Investigation (FPI) \cite{Pollock:2016}, and spacecraft position by Magnetic Ephemeris Coordinates (MEC) \cite{Baker:2016}. This burst-mode interval provides a data rate of \qty{128}{\Hz} for FGM, \qty{8192}{\Hz} for EDP, a \qty{150}{\ms} cadence for ion distributions from FPI, and a \qty{30}{\ms} cadence for MEC. The data is given in the Geocentric Solar Ecliptic (GSE) coordinate system.

We first convert the measurements from the spacecraft frame (primed) to the shock-rest, normal incidence frame (NIF, unprimed), where the shock is stationary and the upstream bulk velocity is normal to the shock, which can be calculated using the plasma parameters in the regions upstream and downstream from the shock. These upstream and downstream parameters, presented in Table~\ref{tab:table1}, are calculated by averaging over 9~s intervals starting at 6:11:45 and 6:13:24, respectively.  Following the procedure outlined in Ref.~\cite{Paschmann:1998}, the shock normal direction \blue{$\vhat{e}_n=(0.78,0.60,0.18)$} \blue{in GSE coordinates} is computed using mixed-mode magnetic coplanarity (their Eq.~10.17) and the shock velocity along that normal \blue{$U'_{sh}=25.1$~km/s} is computed using conservation of mass flux (Eq.~10.29).  We define an orthonormal coordinate system $(\vhat{e}_n,\vhat{e}_1,\vhat{e}_2)$, where $\vhat{e}_1=\vhat{e}_n \times \V{B}'_u/|\vhat{e}_n \times \V{B}'_u|$ and
$\vhat{e}_2=\vhat{e}_n \times \vhat{e}_1/|\vhat{e}_n \times \vhat{e}_1|$.  Transforming from the spacecraft frame to the NIF also requires a \blue{transform velocity} $\V{U}'_{NIF}=-U'_{sh} \vhat{e}_n -\vhat{e}_n\times(\mathbf{U}'_{u}\times\vhat{e}_n)$.  All ion velocity distributions and electromagnetic fields are rotated into the NIF orthonormal coordinate system and Lorentz transformed to the NIF, where the field transformations are  $\mathbf{E} = \mathbf{E}' - \mathbf{U}'_{NIF} \times \mathbf{B}'$ and $\mathbf{B}=\mathbf{B}'$ in the non-relativistic limit \cite{Howes:2014a}.  Note that we must downsample (by averaging) the EDP measurements to the FGM cadence for this Lorentz transformation.  \blue{The resulting \Alfven Mach number of the shock in the NIF is $M_A=5.51$.}

\begin{table}
  \begin{center}
    \caption{Plasma parameters in the spacecraft frame upstream and downstream of the shock. 
    Parallel and perpendicular are relative to $\V{B}'$.}
    \label{tab:table1}
    \sisetup{
      round-mode          = places, 
      round-precision     = 2, 
      per-mode = single-symbol
    }
    \begin{tabular}{|c|S|S|l|} 
    \hline
       \textbf{Quantity}  & \textbf{Upstream} & \textbf{Downstream} & \textbf{Units}\\
      \hline
      $B'_x$         & 3.6743121743202209      & 11.316993474960327  &\unit{\nano\tesla}       \\
      $B'_y$         & -5.7057117223739624     & -17.125839710235596 &\unit{\nano\tesla}       \\
      $B'_z$         & 2.8716121912002563      & 7.5349847078323364  &\unit{\nano\tesla}       \\
      $|\V{B}'|$         & 7.3689804605205218      & 21.866520274329286  &\unit{\nano\tesla}       \\
      $U'_{ix}$      & -365.56745147705078     & -226.29859161376953 & \unit{\km\per\s}    \\
      $U'_{iy}$      & 71.907012939453125      & 173.75478744506836  &\unit{\km\per\s}    \\
      $U'_{iz}$      & 7.5036981105804443      & 40.697417259216309  &\unit{\km\per\s}    \\
      $n_i$         & 11.273509979248047      & 33.102453231811523  &\unit{\per\cubic\cm} \\
      $T_{i\|}$     & 26.678735256195068      & 75.262556076049805  &\unit{\eV}      \\
      $T_{i\perp}$  & 28.444982528686523      & 119.88818550109863  &\unit{\eV}       \\
      $T_i$         & 27.856233596801758      & 105.01297505696614  &\unit{\eV}       \\
      $T_{e\|}$     & 18.45      & 43.26  &\unit{\eV}       \\
      $T_{e\perp}$  & 18.47      & 42.56  &\unit{\eV}       \\
      $T_e$         & 18.46      & 42.79 &\unit{\eV}       \\
      \hline
      $v_A$         & 47.871167871443483      & 82.898332326484692  &\unit{\km\per\s}    \\
      $v_{ti}$      & 73.052067169453210      & 141.83806345107743  &\unit{\km\per\s}    \\
      $d_i$         & 67.819407237540091      & 39.577967594751506  &\unit{\km}      \\
      $\Omega_{ci}$ & 0.70586237511296535     & 2.0945575875775382  &\unit{\Hz}      \\
      $\omega_{pi}$ & 4.4204523487613587      & 7.5747309989344449  &\unit{\kHz}      \\
      $\beta_i$     & 2.3287183967514506      & 2.9274804158267838  &\\      
      $\beta_e$     & 1.8571211559587961      & 0.52288476380327742  & \\
      \hline
    \end{tabular}
  \end{center}
\end{table}

\begin{figure}
\centering
\includegraphics[width=\linewidth]{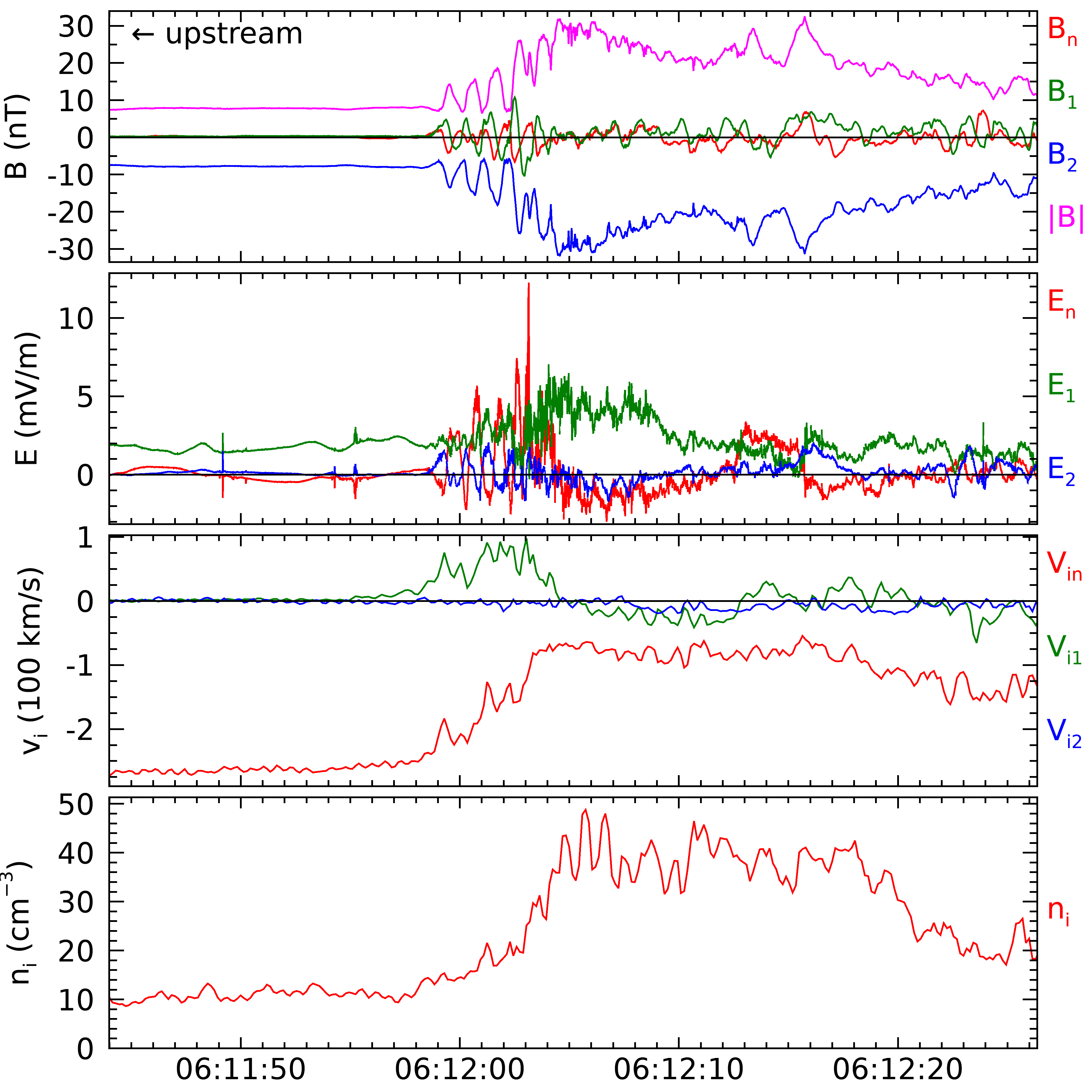}
\caption{Magnetic field, electric field, ion bulk velocity, and ion number density during MMS3's crossing through the shock, rotated and transformed to the shock-rest, normal incidence frame.
}
\label{fig:fields}
\end{figure}

The magnetic field, electric field, ion bulk velocity, and ion density in the NIF from the bow-shock crossing of MMS3 are shown in Figure~\ref{fig:fields}. Note the $\mathbf{B}_u$ is almost entirely in the $\vhat{e}_2$ direction, indicating this is an almost exactly perpendicular shock with $\theta_{Bn} = \qty{90.1}{\degree}$. Within the shock foot and ramp (06:11:58--06:12:04), reflected ions generate a significant $U_{i1}$ component and a non-zero $E_n$ component of the electric field arises from the cross-shock potential.  In this same region, one also observes significant fluctuations of $\V{B}$ with a period of $T \simeq 1.04$~s (used for subsequent boxcar averaging) that likely arise from ion kinetic instabilities in the shock ramp \cite{Burgess:2016,Johlander:2016,Brown:2022}.

To convert the ion velocity distribution function (VDF) measurements from the FPI instrument---measured on a spherical velocity-space grid in GSE coordinates in the spacecraft frame---to the NIF, each velocity-space point is rotated and \blue{transformed} to the NIF, and the measurements at these points are summed into bins in the Cartesian $(v_n,v_1,v_2)$ coordinates with width $\Delta v/v_{ti,u}=0.58$. \blue{The bin width is chosen so that the binning region has 30 bins on each axis and contains 99.9\% of the particles.}  The 3V velocity space may be subsequently reduced to 2V for visualization by integrating along one coordinate, \emph{e.g.}, $f_i(v_n,v_1)=\int dv_2 f_i(v_n,v_1,v_2)$.

Next, we apply the field-particle correlation technique \citep{Klein:2016a,Howes:2017a,Klein:2017b}, using the electric field (downsampled \blue{by averaging} to the FPI cadence of 150~ms) and ion VDF measurements in the NIF, to determine the rate of ion energization as a function of particle velocity $\V{v}$ through the shock transition. Specifically, we compute
\begin{equation}
    C_{E_\lambda}(\V{v},t) = - q_i \frac{v_\lambda^2 }{2} \frac{\partial f_i(\V{v},t)}{\partial v_\lambda} E_\lambda
    \label{eq:fpc}
\end{equation}
where \blue{$E_\lambda$ with $\lambda=n,1,2$ indicates the component of the electric field} in the NIF coordinate system. First, $c'_{E_\lambda}(\V{v},t)= q_i v_\lambda E_\lambda(t) f_i(v_n,v_1,v_2,t)$ is computed at each velocity point, then the $c'_{E_\lambda}$ values are summed into bins in the $(v_n,v_1,v_2)$ coordinates as described above, and finally the velocity derivatives are computed as described in Ref.~\cite{Chen:2019} using centered finite differencing to obtain $C_{E_\lambda}(\V{v},t)$. Note that \blue{integrating the expression in Eq.~\eqref{eq:fpc} over velocity space gives $\int d^3\V{v} C_{E_\lambda}(\V{v},t) = j_{i,\lambda} E_\lambda$, where $j_{i}$ is the ion current density}, which is the rate of electromagnetic work done by the electric field component $E_\lambda$ on the ions.  

\begin{figure*}
\centering
\includegraphics[width=\linewidth]{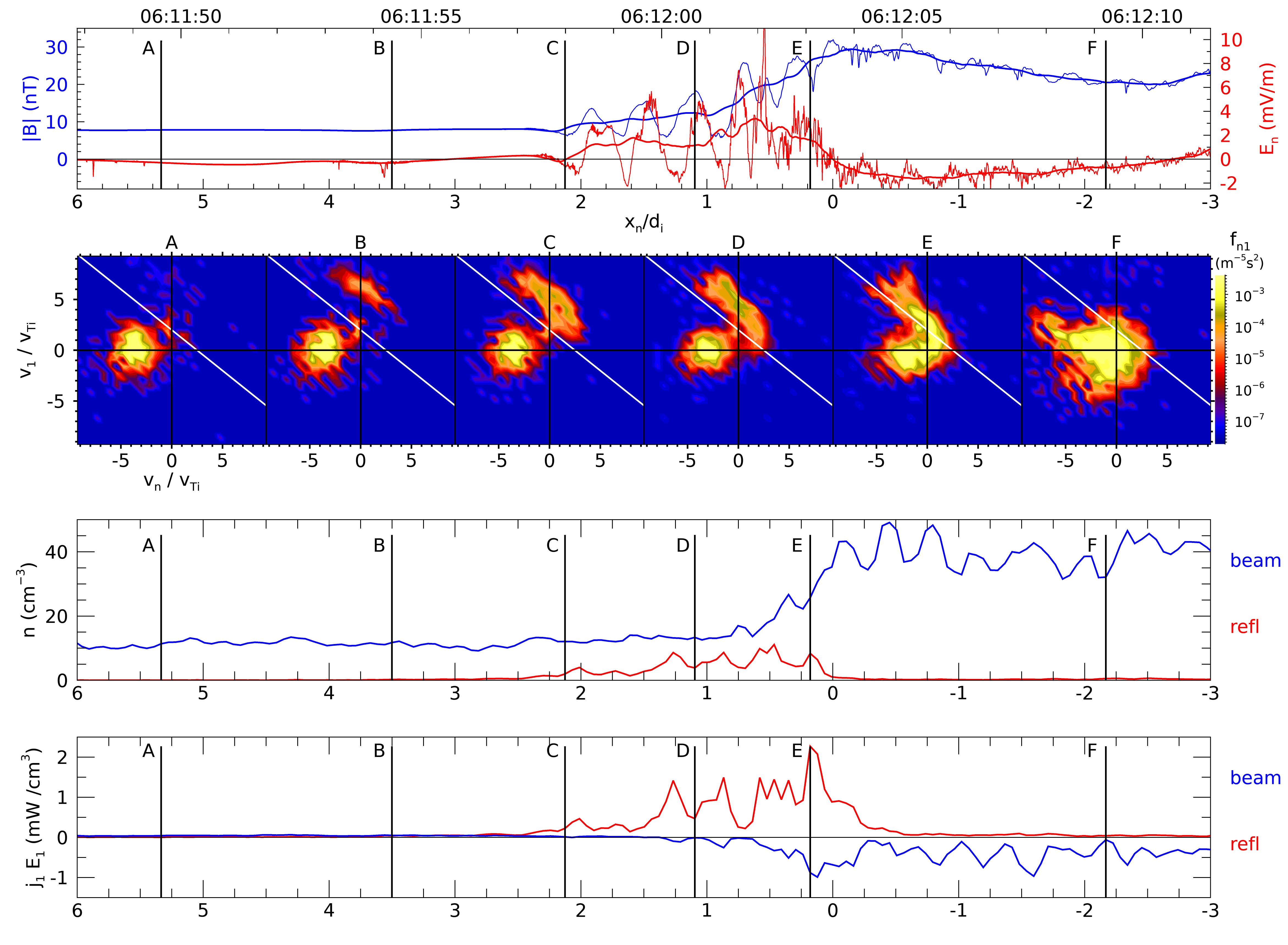}
\vskip -5.05in
\hbox{ \hsize 7.5in (a)  }
\vskip +1.10in
\hbox{ \hsize 7.5in (b)}
\vskip +1.25in
\hbox{ \hsize 7.5in (c) }
\vskip +1.0in
\hbox{ \hsize 7.5in (d) }
\vskip +0.9in
\caption{(a) Magnetic field magnitude $|\V{B}|$ (thin blue) and cross-shock electric field $E_n$ (thin red), along with boxcar-averaged fields $\langle |\V{B}| \rangle_T$ (thick blue) and $\langle E_n \rangle_T$ (thick red), versus the normalized position along the shock normal $x_n/d_{i,u}$. (b) The ion velocity distribution $f_i(v_n,v_1)$ at different positions (A-F) through the shock transition, where the \blue{white} line is used to separate the reflected ions (above) from the incoming ions (below). (c) The number densities of reflected ions $n_{refl}$ (red) and incoming beam ions $n_{beam}$ (blue) vs.~$x_n/d_{i,u}$. (d) Rate of change of ion energy density by the motional electric field $j_{1,i} E_1$  vs.~$x_n/d_{i,u}$, split between the reflected ions (red) and incoming beam ions (blue).}
\label{fig:density}
\end{figure*}

\emph{Results.}---In Figure~\ref{fig:density}a, we present the magnetic field magnitude $|\V{B}|$ (thin blue) and cross-shock electric field $E_n$ (thin red) through the shock, along with boxcar-averaged fields $\langle |\V{B}| \rangle_T$ (thick blue) and $\langle E_n \rangle_T$ (thick red), versus the normalized position along the shock normal $x_n/d_{i,u}$.  Here $x_n=\vhat{e}_n \cdot \V{U}_{s/c}(t) [t-t_0]$, where $\V{U}_{s/c}(t)$ is the spacecraft velocity in the NIF and the shock position $x_n=0$ is defined where the boxcar-averaged cross-shock electric field passes through zero, $\langle E_n (t_0)\rangle_T=0$.  We show logarithmic plots of the reduced ion VDF $f_i(v_n,v_1)$ integrated over $v_2$ in Figure~\ref{fig:density}b at six positions (A--F) through the shock transition. Far in the upstream (A), the ions form a single incoming beam. Approaching the shock (B), a small population of reflected ions at $v_1/v_{ti,u}\simeq 6$ begins to appear \cite{Paschmann:1982,Sckopke:1983,Juno:2021}, and upon reaching the foot of the shock (C), this reflected ion population has grown substantially. As we proceed into the ramp (D), the density of the reflected ion population reaches about one third of the incoming beam density, and \blue{at this position the two populations of incoming and reflected ions are less clearly separated in this projection of velocity space. Near the overshoot (E), the beam and reflected populations begin to overlap in velocity space, forming an irregular, boomerang-shaped distribution. Finally, downstream of the shock (F), the incoming beam has broadened into a wider distribution about the downstream velocity with a significantly higher perpendicular temperature. The reflected population, which underwent shock-drift acceleration (see Figure~\ref{fig:fpc} below) and then passed downstream, can be seen at position (F) as  the small ``knob" of ions at $v_n/v_{ti,u}\simeq -7$ and $v_1/v_{ti,u}\simeq 3$ as they follow their trajectory in velocity space,  orbiting counterclockwise about the downstream $\V{E}\times \V{B}$ velocity \citep{Juno:2021}}.

To explore the acceleration of the reflected ions by the motional electric field $E_1$ through the SDA mechanism, we plot in the first row (a--c) of Figure~\ref{fig:fpc} three views of the ion VDF $f_i(\V{v})$ at the beginning of the foot (C), each integrated over the third velocity component, along with the field-particle correlation with the motional  electric field $C_{E_1}(\V{v})$  \blue{(\emph{c.f.} Eq.~\eqref{eq:fpc})} for the same three views in the second row (d--f). In panel (d), the plot of $C_{E_1}(v_n,v_1)$  shows a blue crescent (loss of phase-space energy density) below a red crescent (gain of phase-space energy density) at $v_1/v_{ti,u} \simeq 5$, corresponding to the acceleration by $E_1$ of reflected ions to higher values of $v_1$, as explained in detail in Juno \emph{et al.} (2021) \cite{Juno:2021}.  Note that integrating over the $v_1$ velocity coordinate in panel (e) shows that the effect of the motional electric field on the ions is a net transfer of energy to the ions (red region).  The first key result of this study is that we have observationally identified this unique \emph{velocity-space signature of shock-drift acceleration} at Earth's bow shock, showing excellent qualitative agreement with the blue-red crescent that was numerically predicted for a perpendicular collisionless shock in Figures~3 and~5 of Ref.~\cite{Juno:2021} (reflected about the $v_1=0$ axis, \blue{as the $B_u$ in the numerical simulations was in the opposite direction to that observed by MMS}). \blue{Note that we find qualitatively and quantitatively similar results for Figure~\ref{fig:fpc} using measurements from the other three MMS spacecraft, as is expected since those spacecraft separations $l_j = 16.9, 16.4, 17.0$~km from MMS3 are significantly less than the upstream ion inertial length, $d_{i,u} = 67.8$~km.}

\begin{figure*}
\centering
\includegraphics[width=\linewidth]{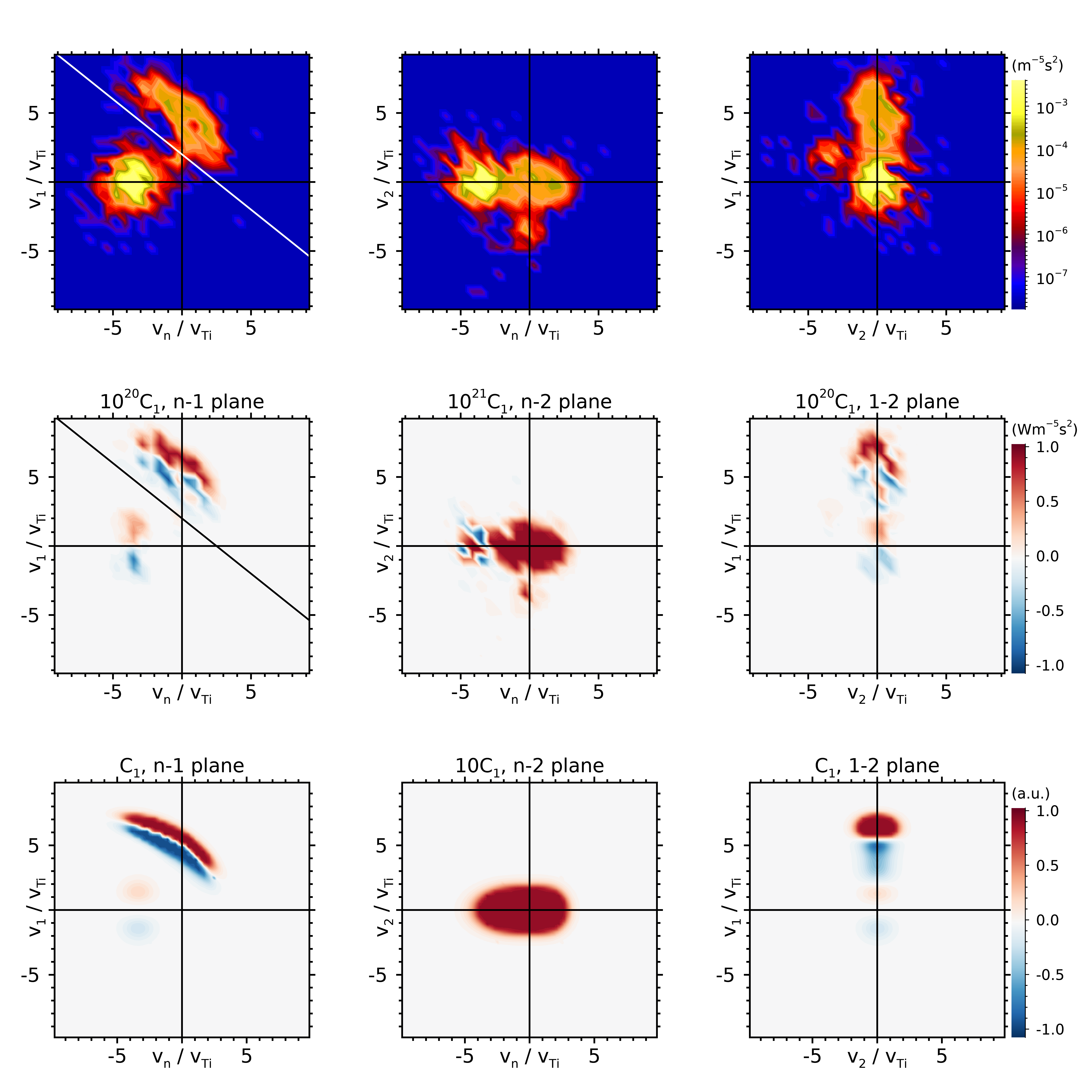}
\vskip -6.75in
\hbox{ \hsize 7.5in (a) \hspace{2.in} (b) \hspace{2.in} (c) }
\vskip +2.20in
\hbox{ \hsize 7.5in (d) \hspace{2.in} (e) \hspace{2.in} (f) }
\vskip +2.20in
\hbox{ \hsize 7.5in (g) \hspace{2.in} (h) \hspace{2.in} (i) }
\vskip +1.90in
\caption{Three views of the reduced ion velocity distribution at the edge of the shock foot (C), showing (a) $f_i(v_n,v_1)$, (b) $f_i(v_n,v_2)$, and \blue{(c)} $f_i(v_2,v_1)$. Comparison of the field-particle correlation with the motional electric field $C_{E_1}(\V{v})$ for the same three reduced views, showing the observational results in the second row (d--f) and the prediction from a Vlasov-mapping model in the third row (g--i).}
\label{fig:fpc}
\end{figure*}

To verify this velocity-space signature of SDA quantitatively, we utilize a Vlasov-mapping technique \cite{Scudder:1986,Kletzing:1994} to map the ion velocity distribution along single-particle-motion trajectories along a piecewise-linear approximation of the boxcar-averaged, observed electromagnetic fields in Figure~\ref{fig:fields}, as detailed in Appendix~D of Ref.~\cite{Juno:2021}. The resulting Vlasov-mapping predictions for the field-particle correlation $C_{E_1}$ are plotted in the third row (g--i) of Figure~\ref{fig:fpc}, showing excellent agreement of the qualitative features in \blue{all three planes, as well as the relative magnitudes among the three views}.  \blue{Similar agreement is found at points (D) and (E), presented in Supplemental Material in Figures~S1 and~S2.}

In addition to using the qualitative features of the velocity-space signature to identify the mechanism of shock-drift acceleration, we can exploit the ability of the field-particle correlation technique to resolve particle energization rates as a function of the velocity to separate quantitatively the rate of energization of different populations of ions.  In Figure~\ref{fig:density}b, we draw a \blue{white} diagonal line \blue{by eye} on the plot of $f_i(v_n,v_1)$ to separate the reflected ion population (above) from the incoming ion beam population (below). In Figure~\ref{fig:density}c, we plot the number densities of the reflected ions $n_{refl}$ (red) and incoming beam ions $n_{beam}$ (blue) vs.~$x_n/d_{i,u}$, found by integrating the VDF over the regions on each side of the \blue{white} line. This shows a significant population of reflected ions in the shock foot and shock ramp (from C through E). We can separate the energization due to the motional electric field $E_1$ by integrating $C_{E_1}$ over the two velocity space regions in Figure~\ref{fig:fpc}d, showing in Figure~\ref{fig:density}d the energization rates of the reflected ions (red) and incoming beam ions (blue) vs.~$x_n/d_{i,u}$.  These plots show that the net ion acceleration by the motional electric field acts entirely upon the reflected population and dominates the ion energization in the foot and ramp.

At position (C) at the edge of the foot,  we find that the beam population has density \qty{12.1}{\per\cubic\cm}, \blue{energization rate \qty{12.9}{\uW\per\cubic\cm}, and energization per ion of \qty{1.07}{\uW}/ion}, while the reflected population has density \qty{1.98}{\per\cubic\cm}, energization rate \qty{225.}{\uW\per\cubic\cm}, and energization per ion of \qty{114.}{\uW}/ion. The shock-drift accelerated ions at this position are thus receiving \blue{107} times more energy \blue{per ion} from the motional electric field than the ions of the \blue{incident beam}. Note that beyond point (E), in the overshoot and downstream regions, \blue{the overlap of the reflected and incoming ion populations in velocity space} prevents the clean separation of their energization rates.

\emph{Conclusion.}---Here we have identified the unique velocity-space signature of shock-drift acceleration for a perpendicular collisionless shock, the blue-red crescent shown in Figure~\ref{fig:fpc}d  first predicted using numerical simulations and analytical modeling by Juno \emph{et al.} 2021 \cite{Juno:2021}, using MMS spacecraft observations for the first time.  The agreement between the signature created using a piecewise-linear approximation of the boxcar-averaged fields in Figure~\ref{fig:fpc}g and the observed signature in Figure~\ref{fig:fpc}d indicates that this unique velocity-space signature is robust to the physics of kinetic instabilities that naturally arise in collisionless shocks \cite{Brown:2022}.  Furthermore, by integrating the field-particle correlation over different regions of velocity space, we can separate quantitatively the energization of the reflected ions, which dominate the ion energization by the motional electric field through the shock-drift acceleration mechanism, from that of the incoming ion beam. 

Having demonstrated here the field-particle correlation analysis of spacecraft observations of a collisionless shock, this paves the way for future use of this technique to characterize fully the energy conversion through a collisionless shock, specifically including the partitioning of upstream bulk kinetic energy into different channels \blue{\citep{Schwartz:2022}}, including bulk heating of the downstream ions and electrons, nonthermal acceleration of a small population of either species, and increase of magnetic energy. The velocity-space signatures resulting from the \blue{field-particle correlation analysis} hold promise to identify the physical mechanisms governing the energy conversion \blue{for more general shock conditions with $\theta_{Bn}<90^\circ$}, including shock-drift acceleration of singly or multiply reflected ions \cite{Juno:2022}, adiabatic heating of electrons \cite{Juno:2021}, or the physics of non-steady energization via shock reformation or kinetic instabilities arising in the shock transition \cite{Brown:2022}.

\begin{acknowledgments}
This research was supported by NASA grants 80NSSC20K1273, 80NSCC18K1366, 80NSSC18K1371, and 80NSSC18K0643
\end{acknowledgments}

\end{document}